\documentclass{emulateapj}

\usepackage{natbib}
\usepackage{graphicx}
\usepackage[usenames,dvipsnames]{color}
\usepackage{afterpage}
\usepackage{amsmath}
\usepackage[letterpaper,top=4cm,bottom=-.5cm,right=.8cm,left=2.2cm]{geometry}

\shorttitle{Solar Mg II h}
\shortauthors{Schmit et al.}

\begin{document}
\title{Observed Variability of the Solar Mg II h Spectral Line}
\author{D. Schmit\altaffilmark{1, }\altaffilmark{2}}
\author{P. Bryans\altaffilmark{3}}

\author{B. De Pontieu\altaffilmark{1}}
\author{S. McIntosh\altaffilmark{3}}
\author{J. Leenaarts\altaffilmark{4}}
\author{M. Carlsson\altaffilmark{5}}
\affil{Lockheed-Martin Solar and Astrophysics Laboratory}
\affil{Bay Area Environmental Research Institute}
\affil{High Altitude Observatory, National Center for Atmospheric Research}
\affil{Institutet f{\"or} solfysik, Stockholms Universitet}
\affil{Institute of Theoretical Astrophysics, University of Oslo}
\begin{abstract}
The Mg II h\&k doublet are two of the primary spectral lines observed by the Sun-pointing Interface Region Imaging Spectrograph (IRIS).
These lines are tracers of the magnetic and thermal environment that spans from the photosphere to the upper chromosphere.
We use a double gaussian model to fit the Mg II h profile for a full-Sun mosaic dataset taken 24-Aug-2014.
We use the ensemble of high-quality profile fits to conduct a statistical study on the variability of the line profile as it relates the magnetic structure, dynamics, and center-to-limb viewing angle.\\
\indent The average internetwork profile contains a deeply reversed core and is weakly asymmetric at h2.
In the internetwork, we find a strong correlation between h3 wavelength and profile asymmetry as well h1 width and h2 width.
The average reversal depth of the h3 core is inversely related to the magnetic field.
Plage and sunspots exhibit many profiles which do not contain a reversal.
These profiles also occur infrequently in the internetwork.
We see indications of magnetically aligned structures in plage and network in statistics associated with the line core, but these structures are not clear or extended in the internetwork.
The center-to-limb variations are compared with predictions of semi-empirical model atmospheres.
We measure a pronounced limb darkening in the line core which is not predicted by the model.
The aim of this work is to provide a comprehensive measurement baseline and preliminary analysis on the observed structure and formation of the Mg II profiles observed by IRIS.
\end{abstract}
\clearpage
\section{Introduction}
The magnetic structure and thermodynamics of the solar chromosphere are open problems in solar physics with broad implications for stellar atmospheres.
The visible spectrum contains several lines (H$\alpha$, Ca II H\&K) which have allowed us to probe the structure and dynamics of the chromosphere using ground based observatories.
The recent launch of the Interface Region Imaging Spectrograph (IRIS, \citealt{depontieu_14}) provides a new dataset observing one of the most important radiators of the chromosphere, the Mg II  h\&k lines at 2803.5\AA~and 2796.4\AA, respectively.
While there have been numerous missions to observe the Mg II h\&k lines the measurements have been sparse.
The earliest measurements were done using a rocket-borne spectrographs \citep{durand_49}.
The first study to derive structural variations in the profiles was \cite{lemaire_73}.
The OSO-8 \citep{artzner_77} and Skylab \citep{doschek_77} missions provided the first orbital datasets.
The Solar Maximum Mission made the first polarization measurements of the Mg line profiles \citep{henze_87}.
The highest spectral resolution measurements of the profile prior to IRIS were made by the HRTS rocket \citep{morrill_01} and the RASOLBA balloon \citep{staath_95} spectrographs.
Center-to-limb measurements were discussed by \cite{gout_74}, \cite{bonnet_81}, and \cite{morrill_08}.
These measurements have been used to construct model atmospheres for a variety of solar structures.
\cite{gout_77,gout_89} modeled the effect of velocity gradients in the internetwork chromosphere on profile asymmetry.
\cite{lemaire_81} synthesized profiles for plage.
Umbral profiles, which were originally identified as uniquely single peaked, have been studied by \cite{kneer_81}, \cite{lites_82}, and \cite{gurman_84}.
Limb observations in Mg II also provided diagnostics of prominences \citep{vial_79}.\\
\indent In addition to the data points provided by solar observations, multiple missions have conducted studies of stellar emission in the Mg II lines.
The shape of Mg II h\&k varies significantly across stellar types: from pure absorption in Altair to single peaked emission in $\epsilon$ Eri \citep{blanco_82, basri_79}.
The differences between the shape of Mg II h\&k are a strong diagnostic of stellar winds and shocks in extended chromospheres like that of $\alpha$ Ori \citep{uitenbroek_98}.
Variability and activity cycles have been detected in the Mg II h\&k lines in other stars \citep{dempsey_96}.
While stellar chromospheres exist in some form across the cooler half of the Main Sequence, there is a complicated relationship with coronae: a chromosphere is a prerequisite to form a corona, but stars with a chromosphere do not need to have a corona \citep{linsky_79}.
The chromosphere-corona link is an important one for solar-terrestrial studies.
Solar irradiance is a driving factor in determining the ion populations of the Earth's upper atmosphere \citep{solomon_05}.
While Mg II h\&k are unlikely to play a significant role in thermospheric photoionization, Mg II irradiance has been shown to be a superior proxy to EUV irradiance over the F10.7 radio index \citep{lean_09, guo_07}.\\
\indent The complex profiles observed in the solar Mg II doublet can be attributed to the lines' complex formation.
A seminal study was presented in \cite{milkey_74}.
The Mg II doublet are resonance lines for a highly abundant element.
As such they are very optically thick at the line cores, which form at relatively low densities under NLTE conditions.
The formation depth of the profile varies significantly with wavelength.
The core is estimated to form between 1-3 Mm above the ($\tau_{500\mathrm{nm}}=1$ ) photosphere.
At $\pm$0.5\AA~the line opacity is reduced such that the formation layer maps near the temperature minimum, 500 km above the photosphere.
\citet{leenaarts_13a,leenaarts_13b} and \citet{pereira_13} used a radiative MHD numerical simulation \citep{gudiksen_11} to forward model the emission of the Mg II lines.
These authors describe in detail the formation mechanisms of the lines, and how line components can be transformed into practical diagnostics of the solar atmosphere.
Our research is complementary to these papers. 
In this paper, we delve into the profiles of Mg II h observed by IRIS.
We fit the line profile with a double gaussian model.
A post processing routine parses the best fit model into a variety of profile statistics.
We analyze a full-Sun mosaic dataset that offers us several million profiles, allowing us to present an overview of how the line components vary over the structurally distinct regions of the Sun.
In particular, we analyze the variation of Mg II h with magnetic structure, dynamics, and viewing angle.
Section 2 presents the observations and data reduction.
Section 3 presents methodology for fitting the profiles and deriving the profile statistics.
Section 4 describes the types of profiles observed and interpretation on the physical sources of variability.
Section 5 is summarizes the results and discusses what future steps need to be taken.
\begin{table}
\centering
\begin{tabular}{c | c}
Structure & Profiles ($\chi^2<1.5$) \\
\hline
Internetwork & 3.8 (3.3)$ \times 10^6$\\
Network & 2.1 (1.4) $\times 10^5$\\
Plage & 1.2 (0.72) $\times 10^5$\\
Sunspot & 3.8 (3.4) $\times 10^4$\\
Filament & 1.4 (1.2) $\times 10^5$\\
\end{tabular}
\caption{}{Number of profiles in each structural region of the 24-Aug-2014 dataset. Profiles with a qualifying $\chi^2$ are listed in parentheses.}
\end{table}
\begin{figure*}
\centering
\includegraphics[width=0.8\textwidth]{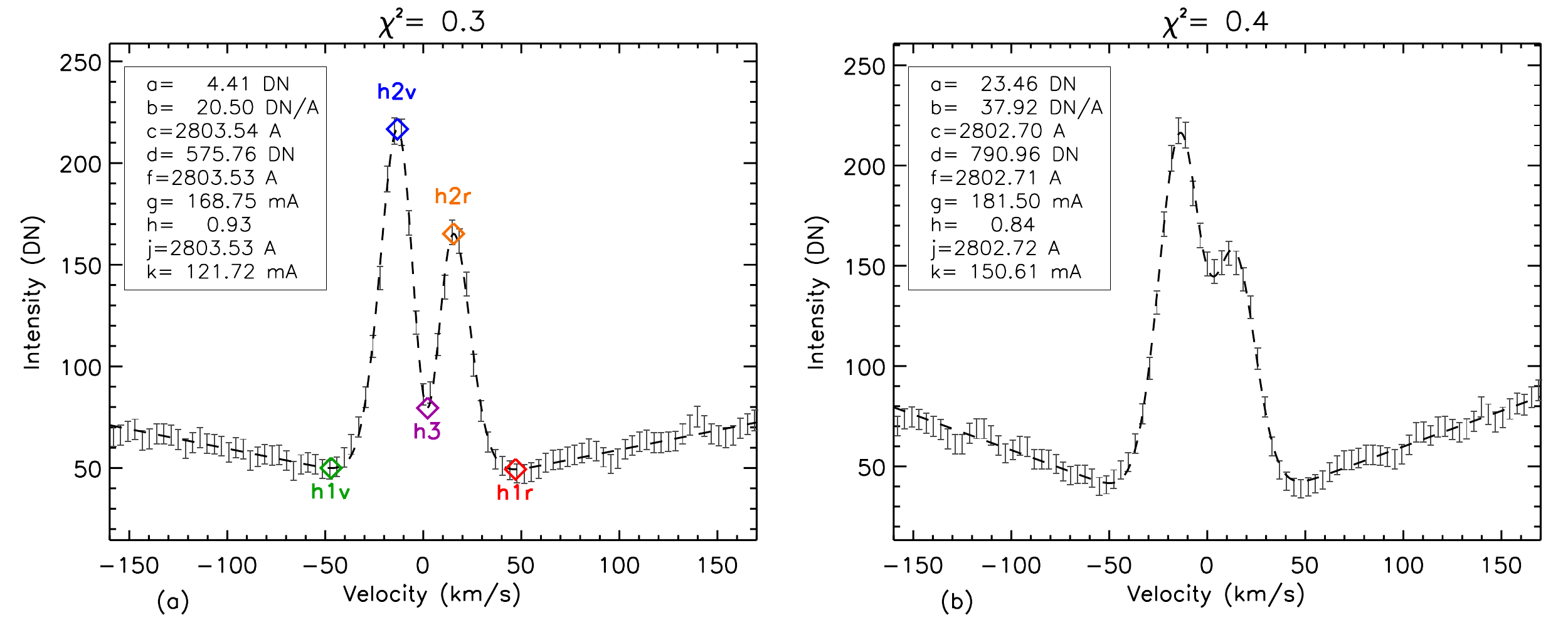}
\caption{Typical double peaked profiles in the internetwork. Grey error bars show the IRIS data. The best fit model is plotted in dashed black. The BFM parameters for Equation 1 are listed in the legend.}
\end{figure*}
\section{Data}
IRIS is designed to investigate the heating of the solar atmosphere by collecting spectra and images of the Sun in three ultraviolet passbands: 1332--1358~\AA, 1381--1407~\AA, and 2783--2834~\AA. IRIS obtains spectral information by passing light from the Sun through a slit and onto a grating. The spatial range of the spectra are limited by the 175~arcsec length of the slit. Temporal limitations are introduced by the exposure time needed to build up sufficient photon counts, and by rastering the slit across the solar disk.\\
\indent We have designed an observing sequence that uses the capabilities of IRIS to build up a spectral map of the entire solar disk over as short a time as practicable. To generate a full-Sun observation with the limited field-of-view of IRIS, we take successive observations at different satellite pointings and build up a mosaic of the Sun. 184 different observations were needed to generate the full-Sun mosaic. Each individual observation consists of a 64-step raster with 2~arcsec steps and 2~s exposure time at each slit position. The spectra along the slit have been binned to resolution of 0.66 arcsec. This gives an area of $128\times 175$~arcsec that takes $\sim 190$~s to observe. All 184 positions takes $\sim 18$~hours, $\sim 10$~hours of which is observing time with the remainder in repointing the spacecraft.\\
\indent The above observing sequence is currently run approximately once per month when IRIS is not in eclipse season. For the purposes of this study, we have chosen one such observation from 2014-08-24 12:16~UT--2014-08-25 05:40~UT. The data used in this paper are IRIS level 2 data products. These data have been processed from the raw observations to remove bad pixels resulting from dust on the detector; dark current and flat field corrections have been implemented; and geometric and wavelength calibrations (based on the rest wavelength of Ni I 2799.17\AA) applied. For a complete description of the calibration process, see the IRIS user guide\footnote{http://iris.lmsal.com/itn26/}. The full-Sun mosaic was assembled by positioning each raster according to its associated spacecraft pointing. There are two sections of the mosaic that were observed while the orbit of IRIS was affected by the South Atlantic Anomaly. Due to an increase in cosmic ray hits at these times, we choose to remove these data from our analysis.\\
\indent This observation contains spectral information for several wavelength windows within the IRIS NUV and FUV passbands. For the purposes of this study, however, we limit our analysis to the Mg II h line at 2803.5~\AA. We analyze a spectral window 3.4 \AA~wide, centered at the Mg II h line, with resolution of $\sim$52 m\AA. 
In future work we intend to expand this analysis to include other emission and absorption features of the IRIS mosaic datasets.\\
\indent In order to compare the IRIS data with the photospheric magnetic field, we use simultaneous observations from the Helioseismic and Magnetic Imager (HMI, \citealt{scherrer_12}) on board the Solar Dynamics Observatory \citep{pesnell_12}. HMI provides the line-of-sight component of the magnetic field for the entire Earth-facing solar disk at a cadence of 720~s. We combine many of these observations to arrive at a full-Sun line-of-sight (LOS) magnetic field map that most closely matches the times of observation of the IRIS data. For each IRIS raster we select the HMI observation that was temporally closest, and then select the portion of the HMI observation that matches the IRIS field-of-view. By repeating this process for all 184 IRIS rasters we can construct an HMI image of the full-Sun.
\section{Method}
We have chosen to use a 9-parameter double gaussian model to fit a 3.4\AA~wide window centered on Mg II h:
\begin{multline}
I(\lambda)=a+b*|\lambda-c|+d*(\exp(\frac{-(\lambda-f)^2}{g^2})\\-h*\exp(\frac{-(\lambda-j)^2}{k^2}))
\end{multline}
where the units of $[I,a,d]$ are DN,  $b$ is DN \AA$^{-1}$, and $[c,f,g,j,k]$ are \AA.
Empirically, we find that this is an accurate model for a majority of the observed profiles.
The Mg II h line is optically thick and the formation height of the profile varies by hundreds of kilometers from h1 to h3.
To calculate a theoretical profile, a synthetic atmosphere must be used to derive the source function as a function of wavelength along the line of sight.
\begin{figure*}
\centering
\includegraphics[width=0.8\textwidth]{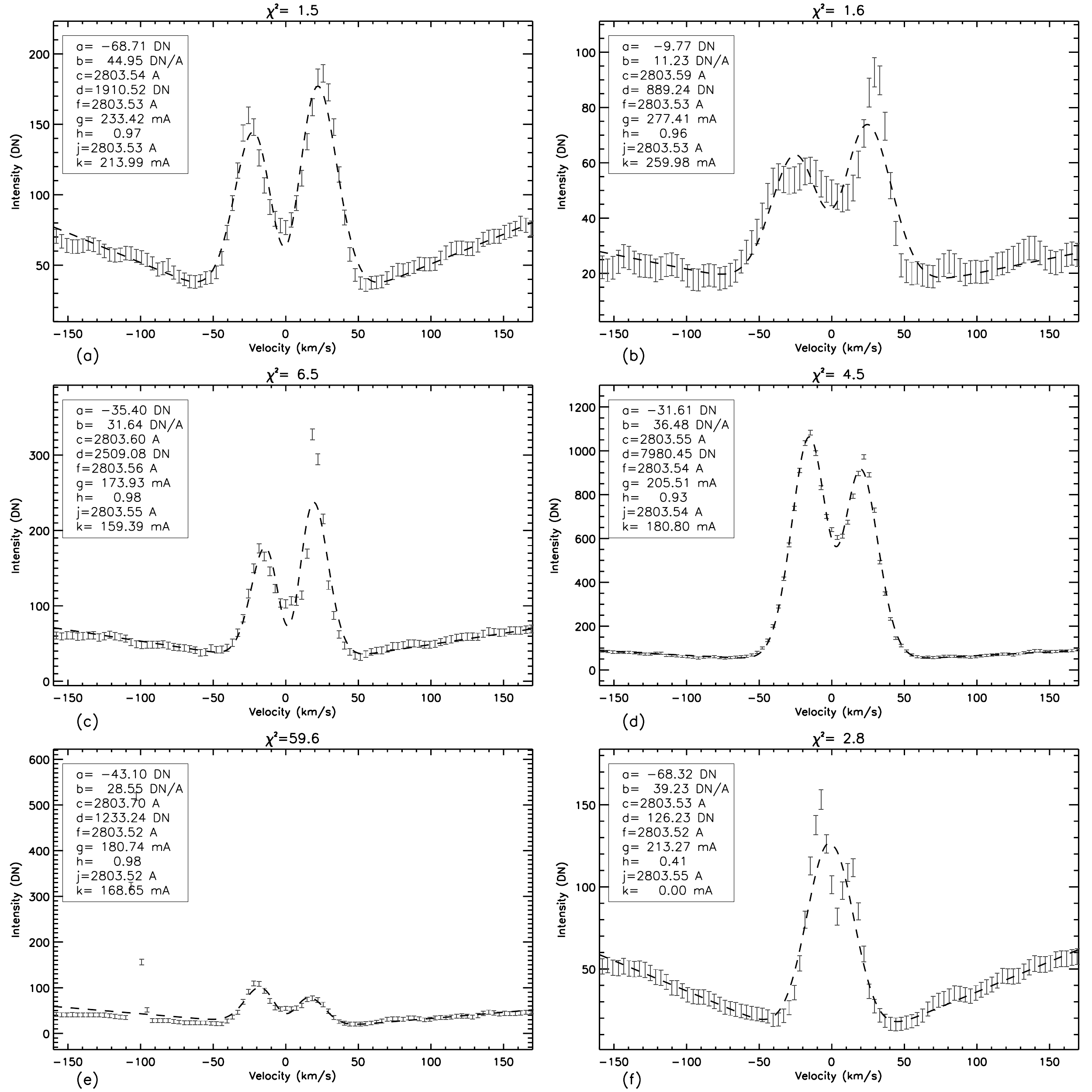}
\caption{Profiles with high $\chi^2$: non-gaussian inner wings (a), flat core (b), non-gaussian asymmetry (c), bright plage (d), cosmic rays (e), faulty minimization (f).}
\end{figure*}
Our fit model does not attempt to include these physics, however the the double gaussian model provides a easily parameterized model that is capable of producing both single and double peaked profiles with incongruous widths in the wing and core.
The average solar Mg II h\&k profile is an emission line with a reversed (depressed) core, which is captured by the superposition of a wide positive amplitude gaussian and a narrow negative amplitude gaussian.
The far wings of the profile are captured by the linear function.
While this technique is applicable to Mg II k as well that analysis is complicated by the presence of the Mn I line at 2795.6\AA.\\
\indent We use the MPFIT least squares minimization algorithm \citep{markwardt_09} to derive a best fit model.
We assume the measurement errors are the linear combination of Poisson noise (based on the 18 photon per DN estimate in \citealt{depontieu_14}) and a constant readout noise of 3 DN.
After a best fit model (BFM, in terms of $\chi^2$) is retrieved from the minimization routine, we apply an algorithm to determine the number and location of local extrema in the BFM.
The average profile will have two maxima at the red and blue edge of the spectral window at (see Figure 1, at 170 km/s and -160 km/s respectively), bounding three minima (h1v, h3, h1r) and two additional maxima (h2v, h2r).
The BFM value at these spectral positions is recorded along with the position.
One additional case is accepted as a viable model: three maxima-two minima (single peak with weak wings).
While some profiles are found with negative value for $b$ in Equation 1, these profiles are the result of noisy spectrums or high total $\chi^2$.\\
\begin{figure*}
\centering
\includegraphics[width=.95\textwidth, trim=0mm 5mm 3mm 8mm,clip]{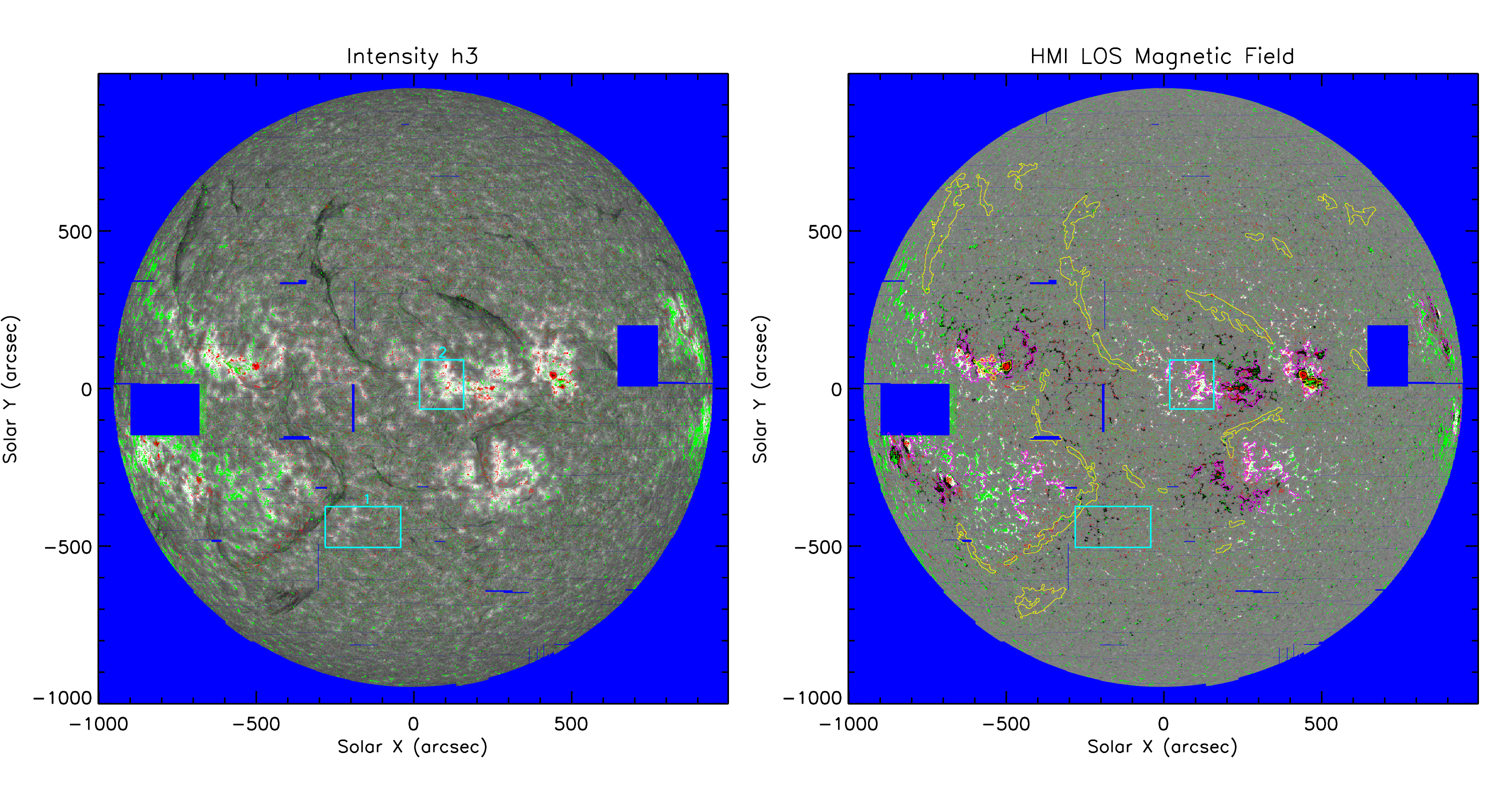}
\caption{Overview of structure on 24-Aug-2014.  Greyscale is the intensity at h3 where a successful fit was retrieved (left). Red pixels are single peak profiles. Green pixels are regions where a successful fit was not achieved. Blue pixels are pixels with insufficient data for fitting. The light blue boxes mark the regions displayed in Figure 4 and 5. SDO/HMI LOS magnetic field map (right). Filaments are bordered by yellow contours and plage is bordered by magenta.}
\end{figure*}
\indent Incorporated with the IRIS package available for Solarsoft is the iris\_get\_mg\_features algorithm which may also be used to identify the h2 peaks and h3 minimum (but not h1 minima) of the h-line.
We have opted for a fitting-based approach to derive a comprehensive parametrization of the profile that uses all the available data.
A low-$\chi^2$ fit represents an accurate model, while a high-$\chi^2$ fit can be further examined to determine how the model has failed.
In a test sample, our fitting method and iris\_get\_mg\_features differed by less than a spectral pixel for $\lambda_{h3}$ for 85\% of profiles.\\
\indent We created structural masks to aid in parsing the dataset.
The internetwork and network subsets were identified using the SDO/HMI LOS magnetic field map. 
The magnetic qualifier for internetwork is that less than 5\% of the pixels (0.5" pixels in the native resolution) within a 5 pixel radius could have an unsigned LOS strength greater than 60 G. 
For network, more than 15\% of pixels need to exceed that magnetic threshold.
Plage was identified as large contiguous regions of unsigned magnetic flux with high intensities at 2803.3\AA~(far enough from into the wing to be unaffected by the highly variable core intensity).
The distinction between large network concentrations and plage was made manually based on the unipolarity of the nearby area.
Sunspots were identified as contiguous regions of high unsigned magnetic flux with low intensities ($<$40 DN) at 2801.8\AA~(the outer most extent of our spectral window and the closest to continuum intensity).
Filaments were identified using only a spectral qualifier.
Filaments were selected as contiguous regions where the line core intensity drops below 38 DN.
Table 1 identifies the number of profiles in each region.\\
\indent Figure 1 shows typical profiles observed by IRIS in the internetwork.
The extrema of the fit are labeled in the standard method: the reversed core is h3, the emission peaks are h2, and minimum in the wing is h1, $v$ and $r$ refer to the violet and red side of the profile.
Our models fits return reduced $\chi^2$ values 0.4 ,which imply that our errors are a liberal estimate of the noise in the spectra.
Approximately 48\% of the on-disk spectra have fits with lower  $\chi^2$ than these models.
We use $\chi^2<1.5$ as our viable model threshold, which applies to 86\% of our dataset.\\
\indent The profiles in Figure 1a and 1b have much in common and simultaneously many differences.
Both profiles are double peaked, but the two peaks are not identical in intensity.
To measure this difference we use the asymmetry statistic:
\begin{displaymath} R_{h}=\frac{I_{h2v}-I_{h2r}}{I_{h2v}+I_{h2r}}
\end{displaymath}
which is also used in \citet{pereira_13}.
Figure 1a and 1b have $R_h$ values of 0.13 and 0.15 respectively.
The h1 width ($\lambda_{h1r}-\lambda_{h1v}$) and h2 width ($\lambda_{h2r}-\lambda_{h2v}$) are also similar for these profiles.
However the core of the profiles are strikingly different.
Both profiles have a h3 minimum, but Figure 1a has a much deeper core depression.
We measure the relative depth of the core using the depth statistic:
\begin{displaymath} D_{h}=1-\frac{2I_{h3}}{I_{h2v}+I_{h2r}}
\end{displaymath}
Figure 1a and 1b have $D_h$ of 0.58 and 0.23, respectively.
The differences of these profiles is emblematic of the complexity we observe in high resolution IRIS Mg II spectra.
The correlations found in \citet{leenaarts_13b} suggest that the intensity at the h3 core is inversely related to the altitude of the $\tau =1$ layer, implying that Figure 1b has less plasma in the upper chromosphere (pushing the $\tau=1$ layer lower) relative to Figure 1a.
The extra chromospheric mass in Figure 1a could be related to dynamic mass loading from shocks or to dense overlying magnetic field.
Further analysis on the properties of the h3 core will be discussed using maps of the h3 intensity in various magnetic structures.\\
\indent Approximately 14\% of the profiles for which we have sufficient data to fit have low quality BFM and are disregarded for statistical analysis.
These points occur throughout the dataset.
There are several distinct types of faulty fits which are displayed in Figure 2.
The most common bad fit look similar to Figure 2a.
We have calculated the contribution to $\chi^2$ from three regions: far wing (window edge to h1), wing (from h1 to h2), and core (h2 to h3).
The error ratio for Figure 2a is approximately 2:7:6 (far wing:wing:core, respectively).
The wing dominates the error as the positive gaussian profile does not accurately capture the changing slope of the profile between h1 and h2.
The model attempts to split the difference so the profile is too wide at h1 and too narrow at h2.
Of the 5$\times 10^5$ bad fits, 46\% are dominated by wing errors.
Figure 2b shows the second most common fitting problem: irregular cores.
Many h3 cores tend to be flatter than the gaussian model.
This is particularly true for low-$\mu$ profiles and inside filaments.
Approximately, 29\% of the bad fit sample is dominated by core errors. \\
\indent In addition to flat cores, the double gaussian model is fundamentally limited in its ability to capture strong h2 asymmetries.
Figure 2c shows one such profile, where the model fits well the blue peak well but poorly the red peak.
We find that $1\%$ of the bad fits have a good fit for one peak and very poor fit for the other.
That percentage is likely higher in actuality but is difficult to statistically identify.\\
\indent Some plage regions contain many bad fits.
Indeed, approximately 45\% of the brightest plage regions ($\max[I]>800$ DN) are bad fits.
Figure 2d presents a typical bad fit plage profile.
While the model captures the overall shape of the profile quite well, the small size of the error bars (primarily based on photon noise) magnifies the inadequacies of the double gaussian model.
The bad plage profiles account for another 2\% of the bad fits in the dataset.\\
\indent
Figure 2e illustrates a cosmic ray hit data.
We estimate that another 2\% of bad fits are the result of anomalous spikes in the spectra.
Figure 2f illustrates a fit where the minimization routine honed in on a single peaked solution, where a double peak solution could also work.
To eliminate this category of profiles, models that are single peaked but overestimate the intensity at the line center are flagged and the minimization routine is reinitialized with an narrower positive gaussian and deeper core.
If the second BFM is still single peaked and overestimates the core intensity, the model is flagged.
These profiles account for another 3\% of bad fits.
\section{Results}
\subsection{Variation of Profiles based on Magnetic Structure}
\indent Figure 3 shows the large scale structure of the Sun in h3 intensity and the magnetic field (zooming into the electronic version of Figure 3 is suggested).
The internetwork is largely faint, although there is significant small scale structure.
Bright emission surrounds magnetic concentrations, network.
Plage regions are very bright in h3.
We do not expect the small flux tube effects of \citet{spruit_76} to apply at the estimated formation height of h3 ($\approx1-3$ Mm).
Rather, the strong emission surrounding magnetic concentrations is related to stronger heating \citep{withbroe_77} in the chromosphere which has been well documented in Ca II H\&K \citep{skumanich_75}.
The dimmest emission occurs in filaments.
The filaments we identify in h3 are similar to those visible in BBSO H$\alpha$ \citep{denker_99}.
Filaments are clear indicators of unusual temperature stratifications in the solar atmosphere ($T<10^4$ K plasma at altitudes of tens of Mm,  \citealt{molowny_99}).
The low temperature filament is likely scattering photons out of the LOS, thus reducing the intensity in a narrow Doppler broadened band surrounding the rest frame line center.\\
\begin{figure*}
\center
\includegraphics[width=.85\textwidth,trim=0 1.7cm 0 0]{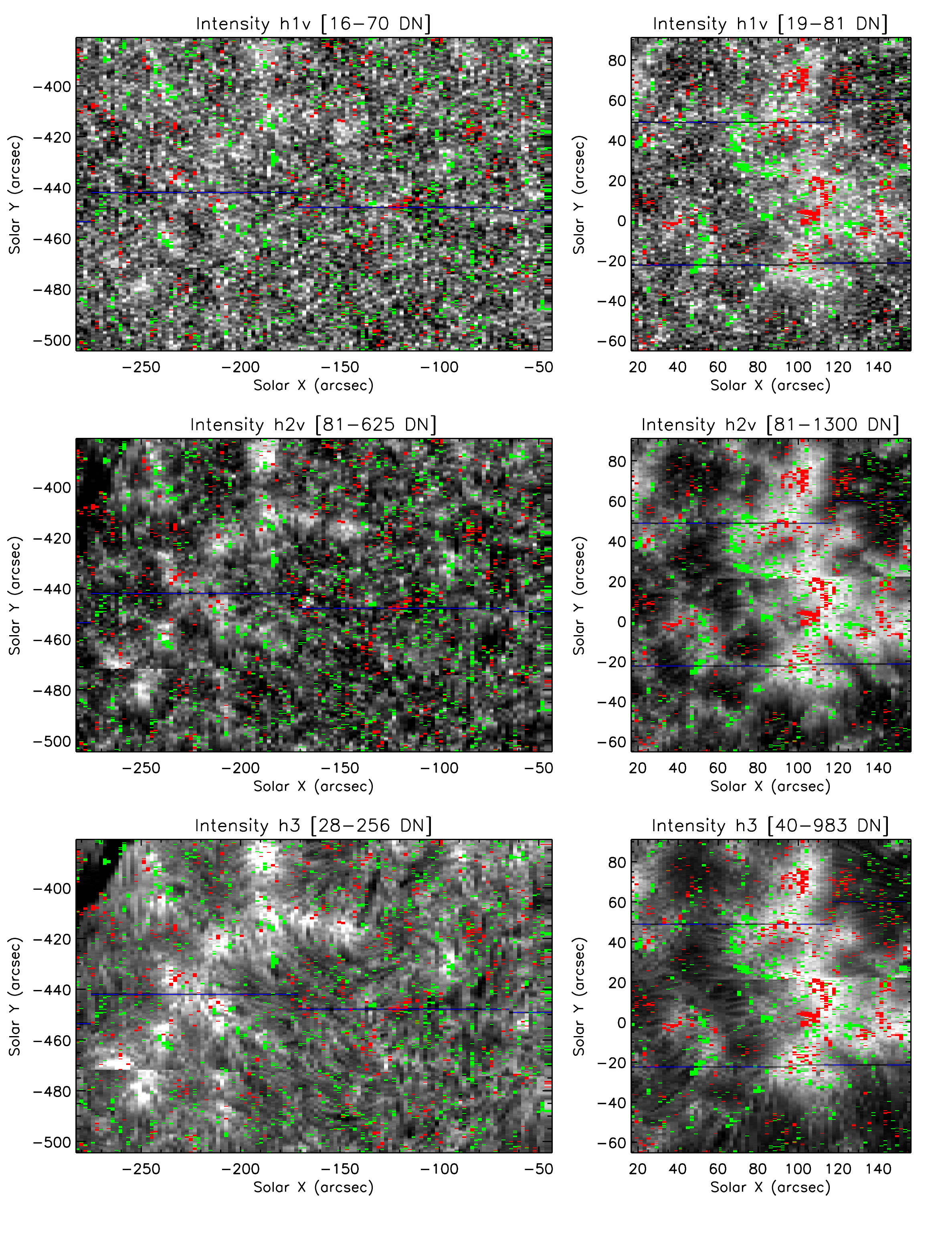}
\caption{Regions extracted from Figure 3 displaying network and internetwork structure (left column) and plage (right column). The color coding is identical to Figure 2. The color table scaling is exponential with $\gamma=0.25$. The colar table bounds are listed in brackets.}
\end{figure*}
\begin{figure*}
\center
\includegraphics[width=.85\textwidth,trim=0 1.7cm 0 0]{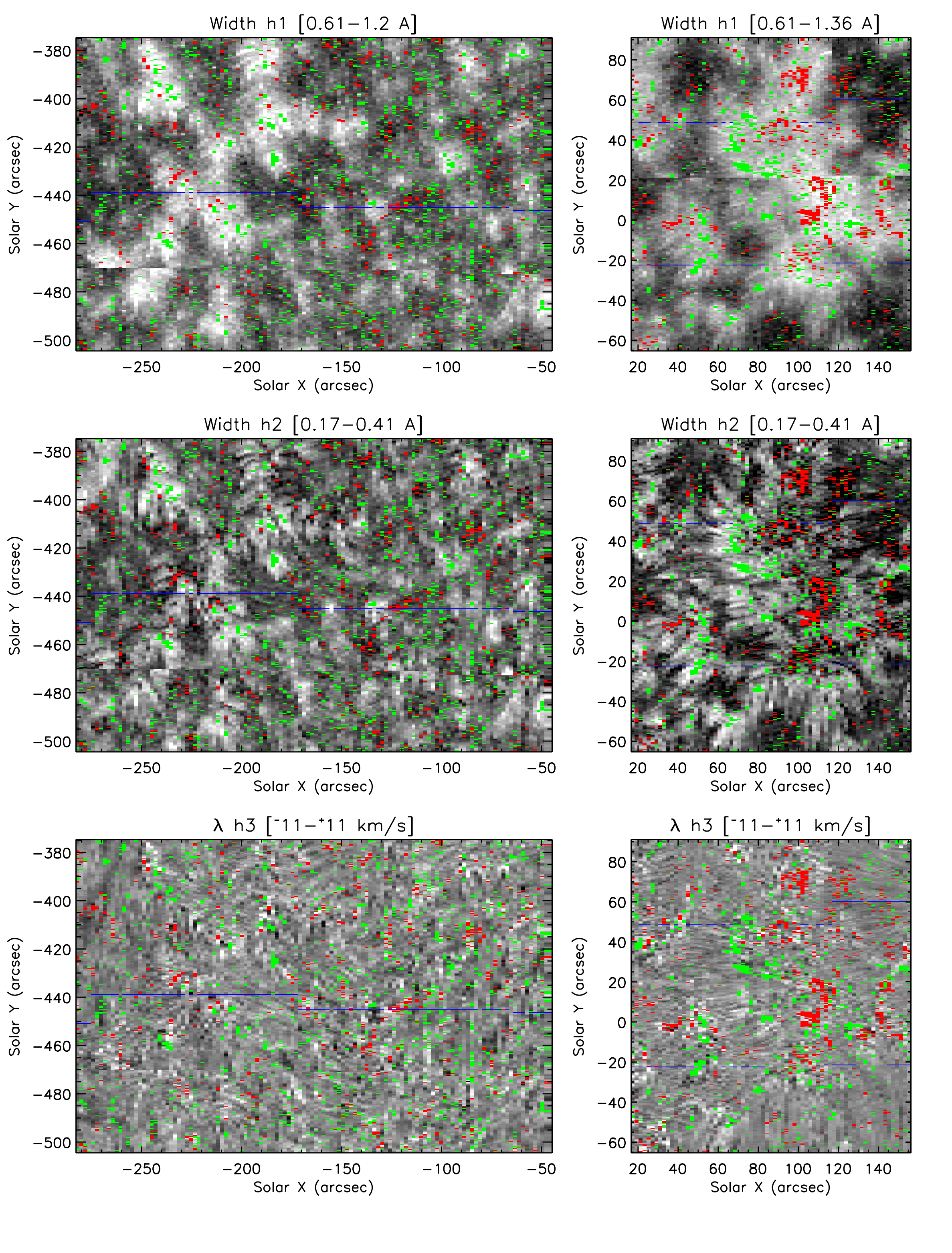}
\caption{Same as Figure 4. The color table scaling is linear}
\end{figure*}
\begin{figure*}
\center
\includegraphics[width=.8\textwidth]{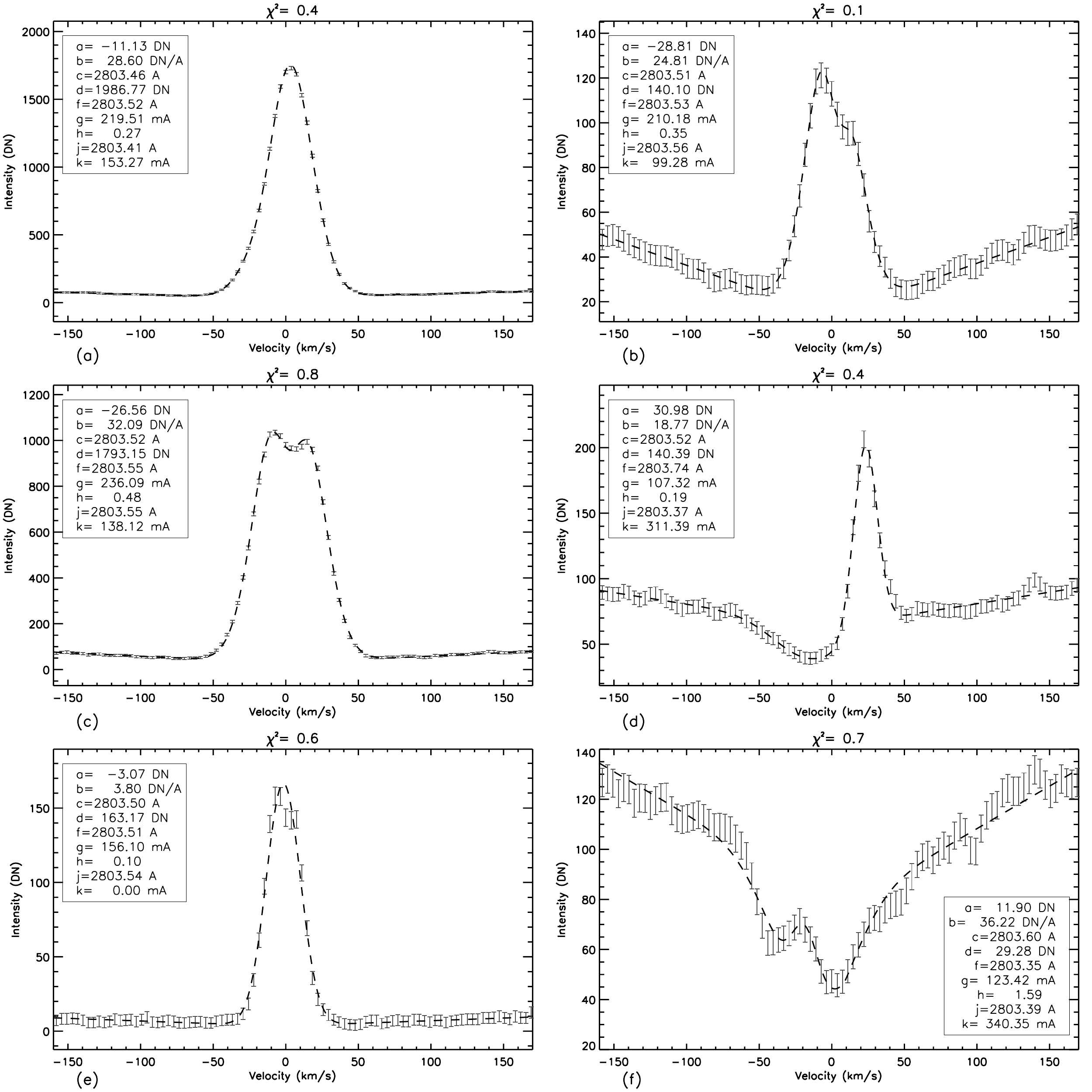}
\caption{ Single peaked profile in plage (a), single peaked profile in internetwork (b), weakly reversed plage profile (c), highly asymmetric single peaked profile (d), sunspot profile (e), reversed absorption profile (f).}
\end{figure*} 
\indent Figure 4 and 5 display two enlarged regions from Figure 3, Region 1 being quiet sun and Region 2 being plage.
We have plotted a variety of statistics in these plots which measure the components of the profiles.
The emission at h1v generally originates in the upper photosphere.
Based on inspection of this and additional higher spatial resolution IRIS datasets, we determine that the structures are a mix of reverse granulation and grains.
At h2v, much of the small granular structure disappears and a diffuse halo of bright structures overlay and surround magnetic concentrations.
The brightest plage is almost twice as bright as the brightest network, and the maximum h2v emission occurs near the center of the magnetic flux concentration.
Single peaked profiles are relativity common ($\sim$10\%) in plage, as depicted by the contiguous red regions in Figure 4.
Figure 6a shows an example of a single peaked plage profile.
While our algorithm separates between single peaked and double peaked profiles to aid in statistical analysis, there is in fact a smooth continuum of weakly separated, weakly reversed profiles that bridge those two profile categories (Figure 6c shows an example).
In the semi-empirical models of the solar atmosphere the Mg II h source function above the temperature minimum has a local maximum at an altitude of 1.2 Mm (see Figure 7 in \citealt{leenaarts_13a}), while the line core forms near a height of 2 Mm.
To produce a single peaked profile the atmosphere must be structured such that source function does not vary significantly between $\tau=1$ at h2 and h3.
At high densities the source function closely adheres to the Planck function, which rises through the chromosphere.
These profiles are consistent with the model that hot high density loops are rooted in plage.
Most sunspot profiles are single peaked, and an example is displayed in Figure 6e.
Single peaked umbral profiles were previously reported in \citet{lites_82} and \cite{morrill_01}.
Sunspot profiles are half as bright at h2 and 20\% narrower at h1 than plage.
The far wing intensity (beyond h1) is lower in sunspots than plage or internetwork.
Single peak profiles also occur in internetwork regions albeit at much lower frequency ($\sim$1\%).
Figure 6b shows a typical internetwork single peak profile.
Although this profile contains a single maximum, it still requires a substantial contribution from the negative gaussian amplitude ($h$ in Equation 1) to achieve a low $\chi^2$.
Figure 6d and 6f show profiles that occur at low frequencies in the internetwork ($<0.0001\%$).
These profiles represent some of the extreme variations from the mean profile shape that we observe in this large dataset.
The profile of Figure 6d could be produced by extreme upflows in the upper chromosphere which completely mask the photons in the nominal position of the h2v peak and wing.
Figure 6f illustrates Mg II h in absorption, which requires that the source function monotonically decrease from the photosphere precluding the mid-chromospheric temperature rise.\\
\indent  The intensity structure at h3 varies significantly in the quiet Sun.
As discussed in \citet{leenaarts_13b}, the Bifrost model predicts that there is a large range of formation heights for h3 (see Figure 4 in that paper).
Semi-empirical models also predict higher column masses at a given altitude in the enhanced network versus faint internetwork \citep{fontenla_09}, which alters the formation height for h3.
The formation region of h3 in the network and in the internetwork likely varies significantly depending on the magnetic topology of the upper chromosphere.
Near the edge of network and extending into internetwork regions, coherent dark loop structures are visible.
These are fibrils and there is a very strong contrast between the h3 intensity in and neighboring the fibril.
Fibrils emanate from a fraction of network boundaries.
In internetwork, it is more difficult to discern large coherent structures.
It is likely that the chromospheric opacity in these regions is lower than in fibrils, and we are able to observe lower in the atmosphere.
The structure at these lower heights is dominated by shocks so we do not observe large coherent horizontal structures.\\
\indent The h2v intensity and the h1 width show similar structures, both exhibit pronounced halos around flux concentrations.
If we consider an atmosphere where the Mg II h profile can be well described by a the linear+single gaussian model (Equation 1 with $h=0$), then we could expect that there would be a strong correlation between the peak intensity and the width of the profile because we are measuring the width based on the inflection point at the continuum and not the gaussian parameter $g$ from Equation 1.
In the solar example where the profiles generally are not well described as a gaussian emission line, we would expect this effect to be reduced.
In the internetwork (Solar-X between -150 and -50 and Solar-Y between -500 and -440) , we see regions of broad h1 width but weak h2v emission which provide a counter example.
The h2 width (also referred to as peak separation in \citealt{leenaarts_13b} and \citealt{pereira_13}) shows significantly more small-scale structure.
The h2 width can be correlated with the range of vertical velocities present in the upper chromosphere \citep{leenaarts_13b}.
The internetwork h2 widths are not much narrower than in network or plage, despite a large difference in $I_{h2v}$.
In plage and network, collimated and linear structures are visible hinting at magnetically aligned features.
The internetwork is likely traversed by complex web of canopy fields connecting many dispersed and mixed-strength magnetic concentrations \citep{schrijver_03}.
Flows along topologically distinct loops that cross through the LOS would not be correlated and might exhibit large velocity gradients that produce wide h2 separations.
Concentrations of low h2 width can be seen surrounding single peaked profiles in internetwork, network, and plage.
As mentioned in the discussion of Figure 6, these profiles fit the gradual transition from weakly reversed double peak profiles to single peaked profiles.\\
\indent The Doppler shift of h3, $\lambda_{h3}$, rarely exceeds 11 km/s.
Near the edge of plage, fibrils show coherent velocity structure.
In the network, it is difficult to discern any fibrils in $\lambda_{h3}$.
Network actually exhibits a greater variability in h3 velocity than plage or internetwork.
The h3 velocity is highly correlated with the vertical velocity at the $\tau=1$ layer.
The plage flow structures are likely related to the periodic flows and transition region oscillations reported in \citet{depontieu_03}.\\
\begin{figure*}
\includegraphics[width=\textwidth]{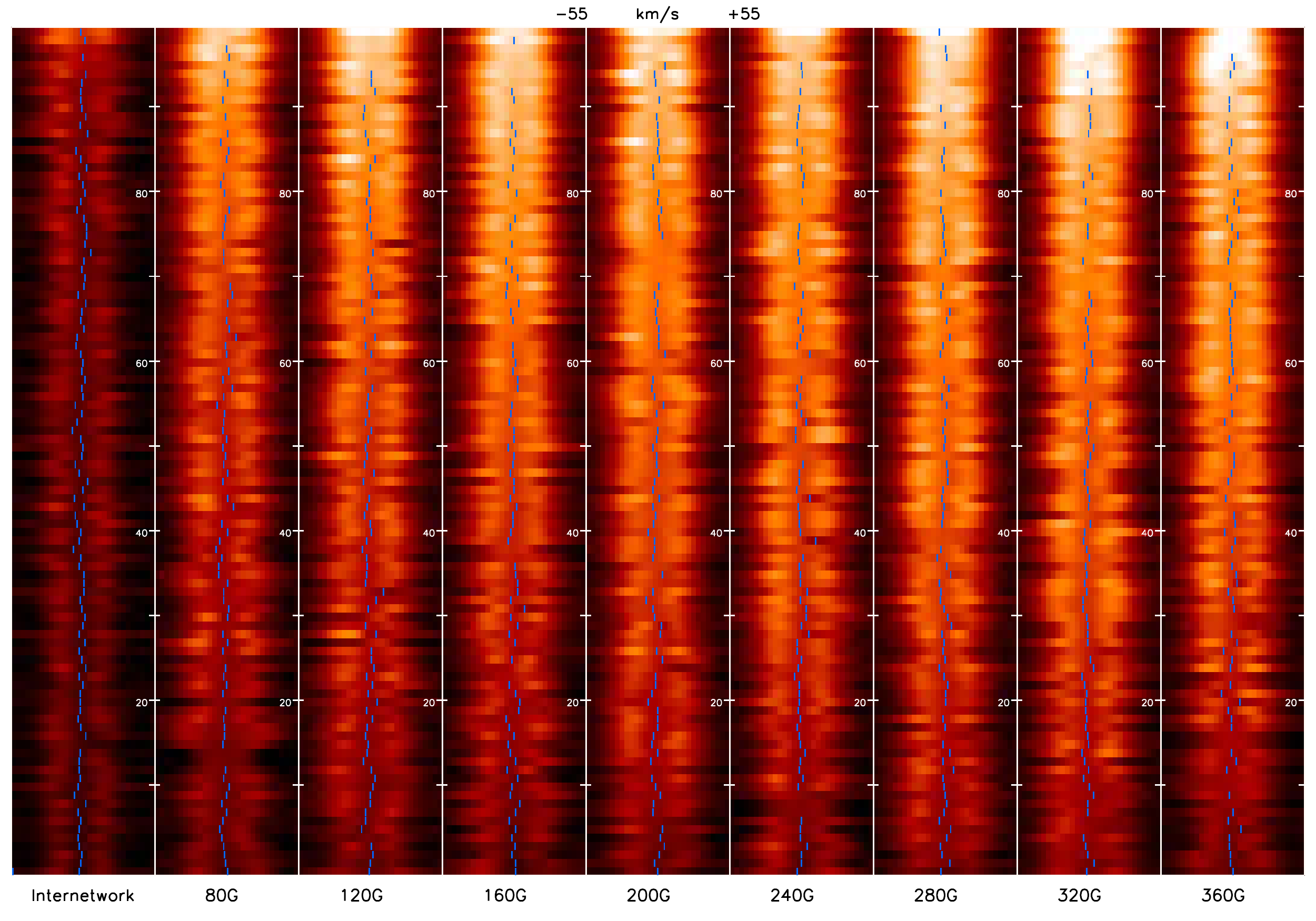}
\caption{Example low-$\chi^2$ profiles from near disk center. On the y-axis are 100 stacked profiles sorted by the intensity at 2803.5\AA~(v=0 km/s). Each x-column represents profiles from regions defined by HMI LOS field strength. The spectral width of each column is $\pm 55$ km/s. The color table is exponential with $\gamma=0.25$ and represents the spectral intensity. Each profile with a well-defined h3 minimum has it marked with a blue bar.}
\end{figure*} 
\subsection{Dynamic Profile Variations}
There are profound differences in the shape of the Mg II h profile and the morphology of spatial structure based on the magnetic environment.
We also know the chromosphere is buffeted by waves driven by the turbulent velocity field of the photosphere.
Many wave modes impart a direct modulation of spectral profiles based on the density and velocity perturbation.
In addition, these waves may add to magnetic stressing or heating, which in turn varies the thermal structure of the chromosphere on small scales.
Based on these effects, we observe a high degree of variability of profiles within the magnetically classified regions. \\
\indent Figure 7 displays the mixture we see of magnetic and non-magnetic profile variability (zooming with electronic version of Figure 7 is suggested).
Each column is sorted from lowest to highest intensity at the rest wavelength line core.
Overall, there is a general trend that the brightest profiles occur in stronger magnetic field, which is consistent with the model that chromospheric heating may be induced or at least aided by magnetic fields.
The brightest profiles also tend to be single peaked.
The intensity tends to be higher at h2v than h2r.
This is an indication that downflows predominate the chromosphere, which is well explained for shocks propagating through a non-magnetized atmosphere in \cite{carlsson_97}.
This effect has also previously been measured in optically thin chromospheric emission lines by \citet{hardi_99}. \\
\indent The comparison between Figure 4 and Figure 7 is interesting.
When structural information is included with the data (i.e. presenting in 2D maps), it is relatively straightforward to categorize how magnetic profiles differs from non-magnetic profiles.
By reducing the dimensionality of the dataset (i.e. presenting the profiles categorized only by magnetic field strength), the inherent variability of the profiles becomes more obvious.
Nearly all the types of profiles that we observe in the internetwork also occur in strong magnetic field regions, although the fractional occurrence rate may change.
Table 2 presents examples of profiles with particular characteristics that are visible in Figure 7.
The examples are selected to span both the magnetic field strength as well as profile core intensity.\\
\indent The variability depicted in Figure 7 is attributable to dynamics as waves and shocks modulate the intensities and shape of the profiles.
These processes should have a highly patterned and repeated spectral signature, similar to the H$\alpha$ sawtooth pattern attributed to passage of a shock in \citet{hansteen_06}.
While our dataset is ill suited for determining those patterns, the temporal signature of the dynamics should be engrained in the distribution of the profiles.
In Section 4.3, we present measurements on the mean and variance of a number of profile statistics.
In addition to averaging the effect of dynamics into a single archetypal profile, it also important to consider the coordinated effects these dynamics have on the shape of the profile.
To extract these relationships from our dataset, we calculate a cross correlation matrix of the profile components, presented in Table 3.
The correlation coefficients allow us to measure if the variability of a particular statistic occurs synchronously with any other statistic.
Here the cross correlation of statistics $S_a$ and $S_b$ is defined as:
\begin{displaymath}
C_{ab}=\sum_{ij} \frac{(S_a^{ij}-\overline{S_a})(S_b^{ij}-\overline{S_b})}{\sigma_a \sigma_b}
\end{displaymath}
where $S^{xy}_n$ where represents statistic $n$ at spatial position $xy$, $\overline{S_n}$ is the mean and $\sigma_n$ is the variance of statistic $n$.
$C_{ab}$ varies between -1 (anti-correlation) and 1 (correlation).
We have limited our profile subset to high-quality double peaked fits of near disk center internetwork so as to limit the scope of variability to fine structure and dynamics and not viewing angle or magnetic structure.
The variables which are most well correlated are the h3 wavelength and and h2 asymmetry.
Because the far wings of the profile extend deeper into the atmosphere and do not shift in wavelength dramatically due to the low sound speed, this correlation is expected based on our double gaussian model.
The more misaligned the negative gaussian is from the positive gaussian, the larger the h2 asymmetry.
Given the success rate of our fitting algorithm and the quality of the fits, we believe this effect is robust against the bias of the fit model.
\begin{table}
\begin{tabular}{c|l}
Profile Characteristic & Examples (Column and Rank)\\
\hline
Unshifted Single Peak & IN-86, 120G-99, 360G-8, 360G-99\\
Shifted Single Peak & IN-15, 360G-30\\
Wide h1 & IN-28, IN-95, 200G-33, 320G-89\\
Narrow h1 & IN-3, 240G-19, 240G-82\\
Wide h2 & IN-6, 200G-63, 240G-65\\
Narrow h2 & 80G-28, 80G-97, 160G-25, 360G-96\\
Asymmetric ($I_{h2v}>I_{h2r}$)& IN-43, 120G-28, 200G-86, 360G-75\\
Asymmetric ($I_{h2r}<I_{h3r}$)& 80G-55, 320G-14, 320G-72\\
Shifted h3 & IN-73, 80G-55, 120G-34, 360G-77\\
\end{tabular}
\caption{}{Notables profiles in Figure 7. Coordinates label the magnetic field and rank.}
\end{table}
The h1 width and the h2 width are also strongly correlated.
Based on a scaling law argument, \cite{ayres_79} suggests that the ratio of h1 and h2 is determined by the vertical extent of the chromosphere and rate of chromospheric heating.
We expect that deeper profiles (high $D_h$) would produce wider h2 separation and thus have a strong correlation.
The relatively high correlation between h2 asymmetry and h2v intensity is an artifact.
By comparing asymmetry with h2r intensity and h-line radiance, we find brighter profiles are generally more symmetric.
The preponderance of positive asymmetry profiles biases the sample.\\
\indent These correlations are best used in conjunction with rapid cadence raster scans that allow us to extract both spatial and temporal information on the evolution of profiles across well resolved spatial structures.
In particular, we need to further develop studies which link the variable chromospheric profiles with the underlying photospheric drivers.
\begin{table}
\includegraphics[width=.5\textwidth,trim=1.8cm 10.5cm 9cm 2cm,clip]{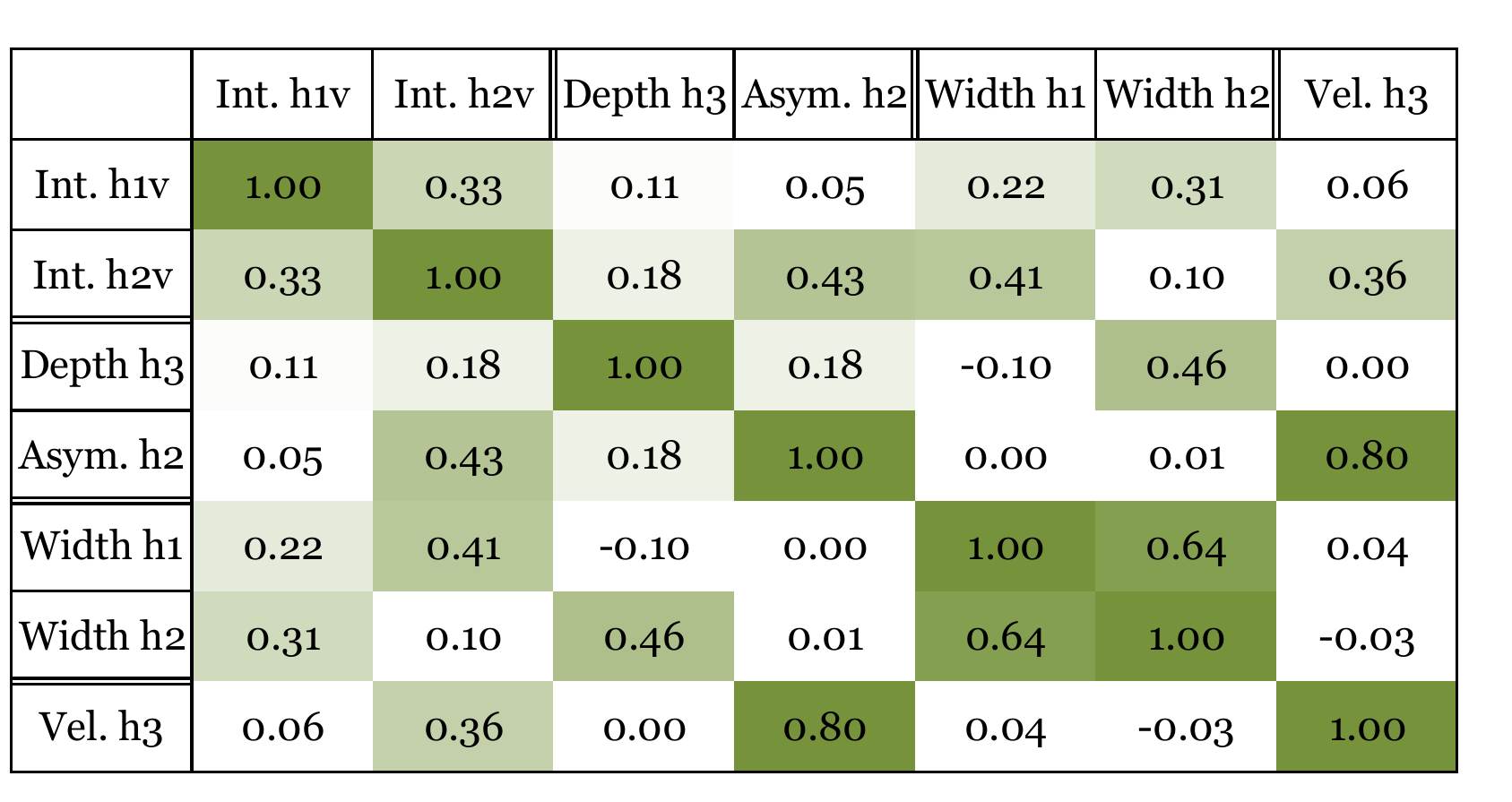}
\caption{}{Cross correlation coefficients for components of internetwork profiles. The ensemble of profiles only uses models with $\chi^2 <1.5$ and $\mu >0.75$. Abbreviations: Int.-intensity, Asym.-asymmetry, Vel.-velocity.}
\end{table}
\subsection{Center-to-Limb Variations}
The viewing angle changes the observed intensities and shape of the profile because the column densities and velocity projections are dependent on the line of sight.
Figure 8 and 9 illustrate the how the viewing angle affects the profiles in our three well populated magnetic subsets using joint probability distribution plots.
The internetwork is by far the largest subset and is evenly distributed in $\mu$.
The center-to-limb trends (slopes) are negative in h1, h2, and h3 intensity (similar to optical continuum) and positive in h1 and h2 widths.
The h2 and h3 intensity distribution are more dispersed at center than limb.
This may be revealing that atmospheric thermal anisotropies decrease as a function of altitude.
As constructed, there is no way to account for the spread of these statistics (at a given angle) using the semi-empirical models.
The asymmetry and depth statistics are unaffected by varying $\mu$.\\
\begin{figure*}[p]
\center
\includegraphics[width=0.85\textwidth]{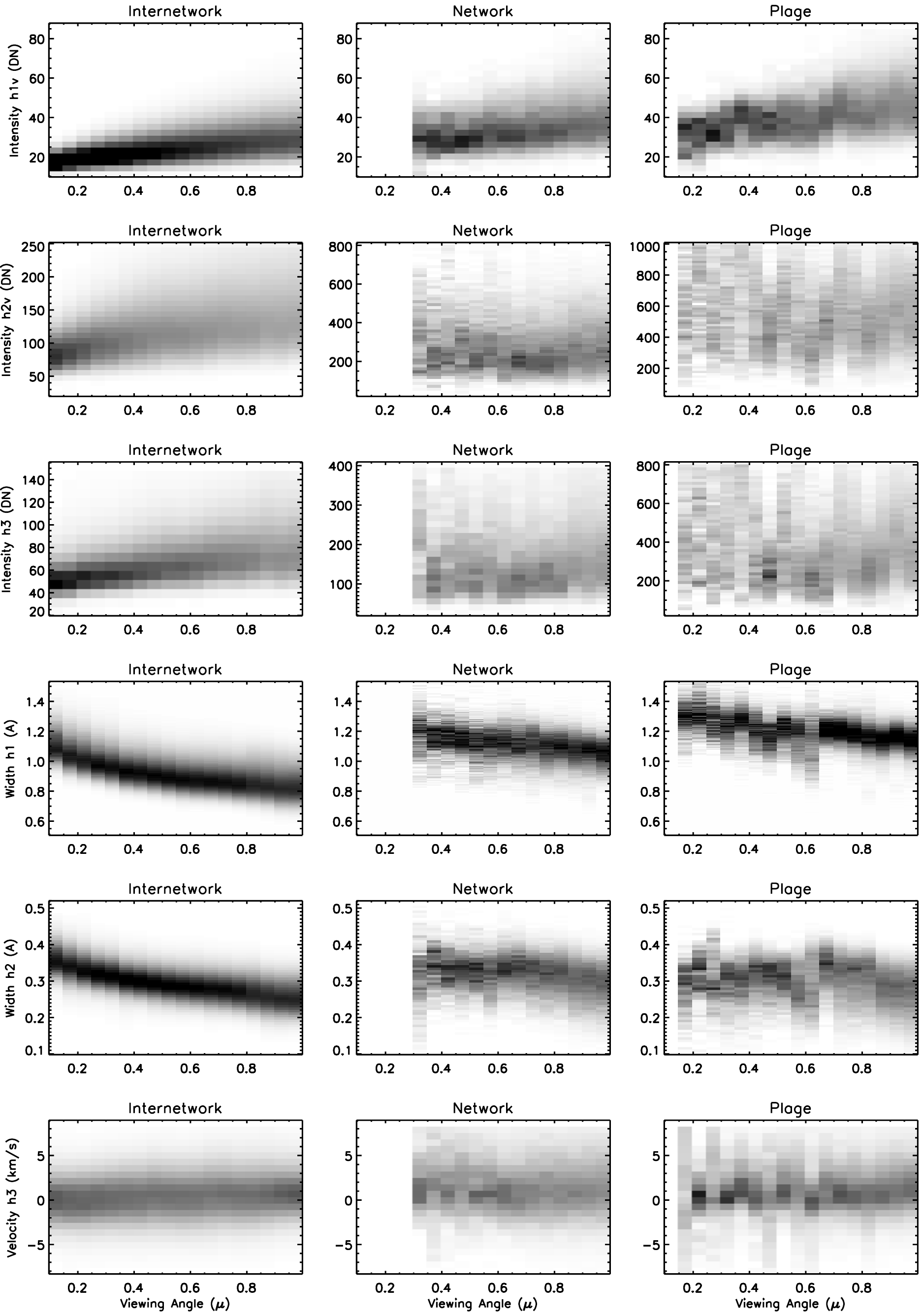}
\caption{Joint probability distributions for the the effect of viewing angle,$~\mu$, on components of the Mg II h profiles. The scaling has been linearly normalized based on bin size and number of applicable profiles at each viewing angle. The maximum occurrence rate is uniform across structures but varies per statistic.}
\end{figure*}
\begin{figure*}
\center
\includegraphics[width=.8\textwidth]{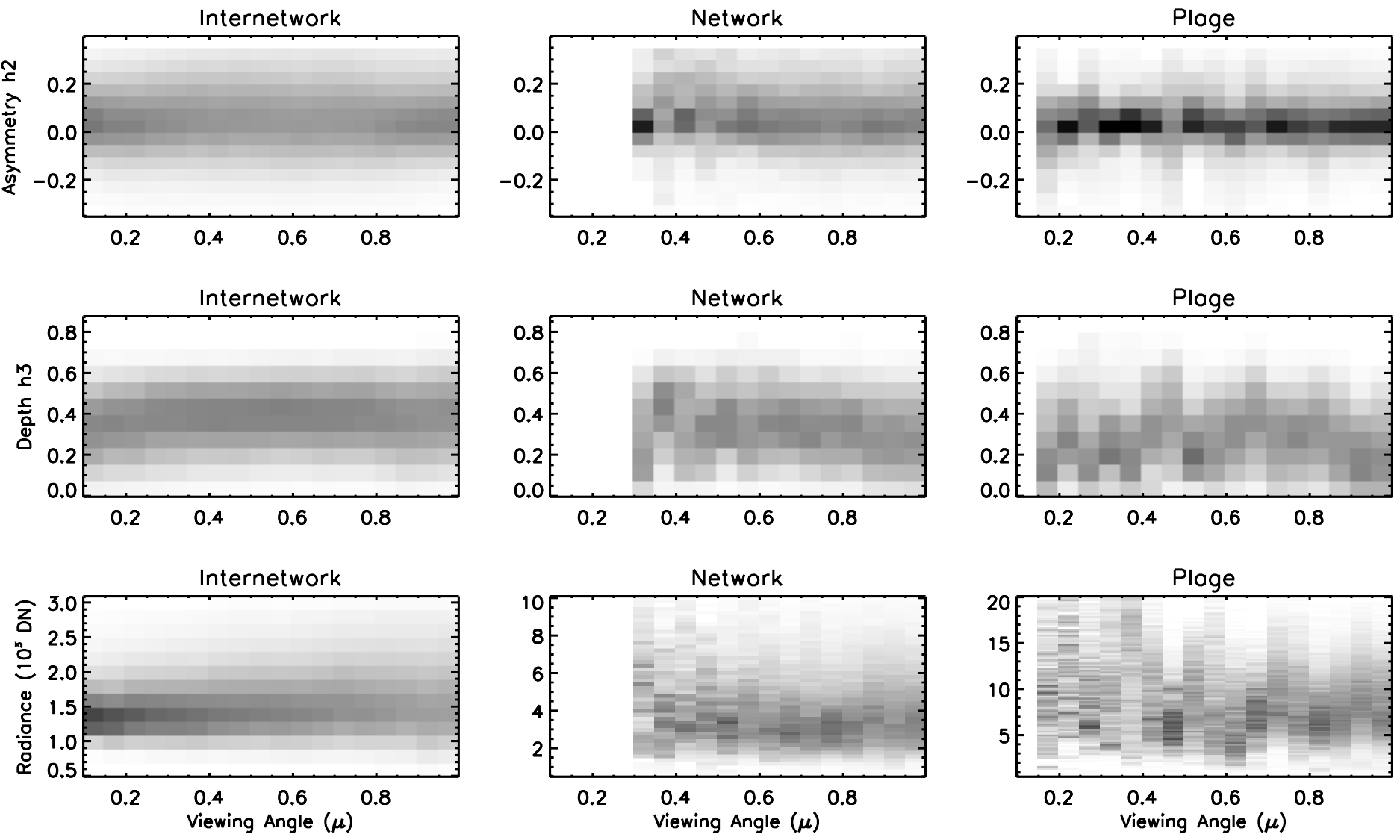}
\caption{Same as Figure 8.}
\end{figure*}
\begin{table*}
\center
\begin{tabular}{c||c|c|c|c}
Statistic & ALM$^1$ Model B & ALM$^1$ Model F & IRIS$^2$ IN & IRIS$^2$ NW\\
\hline
Peak Intensity, $\mu=1.0$ & 210 & 390 & 170 & 280\\
Peak Intensity,  $\mu=0.1$ & 140 & 270 & 90 & -\\
Core Intensity, $\mu=1.0$ & 40 & 130 & 90 & 200\\
Core Intensity,  $\mu=0.1$ & 40 & 120 & 50 & -\\
Depth, $\mu=1.0$ &  0.81 & 0.67 & 0.47 & 0.29\\
Depth, $\mu=0.1$ &  0.71 & 0.56 & 0.44 & -\\
\end{tabular}
\caption{}Comparison of center-to-limb variation from \citet[ALM]{avrett_13} for Mg II k and the measured quantities with IRIS for Mg II h in internetwork (IN and Model B) and network (NW and Model F). $^1$ALM intensity units are W m$^{-2}$ sr$^{-1}$\AA$^{-1}$. $^2$ IRIS intensity units of DN.
\end{table*}
\indent Given the clustered distribution of magnetic flux, the distribution of plage and network fields are less uniform statistical samples.
The magnetic field strength in plage varies significantly from region to region and plays a role in the scatter of the $I_{h2v}$ and $I_{h3}$ distributions.
Plage and network are brighter than internetwork and wider at h1.
For nearly all statistics, the distributions are more dispersed than internetwork.
The h1 widths are well correlated with viewing angle.
The h2 widths have a wider spread with an extended low-width tail.  
Plage and network are less asymmetric and have shallower cores compared to the internetwork.
In order to measure radiance in the line, we have summed the data between h1v and h1r.
In radiance near disk center, network is a factor of 3 brighter than internetwork, while plage is a factor of 8 brighter.
Network is more dispersed in h3 velocity than internetwork or plage.
Given the strong flows often observed in active region coronal spectra, it is surprising that network exhibits a higher percentage of high-velocity profiles than plage.\\
\indent \cite{avrett_13} discussed the center-to-limb variation of the Mg II k profiles based on the predictions of a series of semi-empirical models of the solar atmosphere.
In that paper, Model B represents the mean internetwork while Model F represents the bright network.
Table 4 compares the spectral intensities of the model and our derived profile statistics.
While the semi-empirical models capture the variation of the emission peak (between internetwork and network and center-to-limb), it does not capture the variation in the core.
The semi-empirical model predicts relatively little variation in the core intensity as a function of $\mu$, while we observe a measurable decline in the core at the limb.
As discussed in section 4.1, if the formation height of h3 is highly variable then the limb darkening effect can be explained by the increased sampling of high-lying and dark fibrils at low-$\mu$. \\
\indent The semi-empirical models are constructed as a tool to incorporate a broad range of spectral information and convert it into a consistent model of the atmosphere.
These static models provide us information on the structure of an averaged solar atmosphere, while the 3D MHD models provide information on the dynamics based on physics but limited by the computational power.
By comparing the IRIS data with these models, we increase our understanding of this complex dataset and identify portions of the model which require adaption.
\section{Conclusions}
Most of our understanding of the chromosphere is based on ground based observations of Ca II and H$\alpha$.
While datasets like that of SST/CRISP provide us the highest spatial resolution measurements of the chromosphere, they are limited in spatial FOV and the often brief periods of prime atmospheric seeing.
The IRIS spacecraft has opened up a new window into the chromosphere, with the ability to resolve sub-arcsec structure and the stability and atmosphere-free seeing to produce long-duration datasets.
One of the primary purposes of the analysis of solar spectral lines is to diagnose the physical processes and the thermodynamic conditions of the atmosphere.
Prior to establishing diagnostics, it is essential to identify the methodology for measuring the necessary spectral features, and our works fits into this first step.
The Mg II k line differs slightly from Mg II h in formation height due to the difference in oscillator strength.
It is important to extend this analysis to that line for additional diagnostic capabilities, however our method requires adaption.
The Mn I line at 2975.5\AA~can blend with the k-line for wide profiles.
A well constrained fit model can likely be constructed with a spectral window that also includes the unblended Mn I 2999.1\AA~line.
This is a likely direction for future analysis.\\
\indent \citet{leenaarts_13a,leenaarts_13b} and \citet{pereira_13} have used an advanced numerical simulation of the solar atmosphere to forward model Mg II h\&k emission and compare the derived spectral information which the physical properties of the emitting plasma.
Many of the statistics we have measured in this dataset are identical to those discussed by those authors.
A direct comparison between the modeled profiles and observed profiles is difficult.
While numerical models have proven powerful in describing specific physical processes (the formation of shocks by high frequency vertical velocity fluctuations described the \citealt{carlsson_97}, for example), the models are not producing many of the spectral features we identify in the observations.
We know that the observed profiles are significantly broader than those forward modeled by Bifrost.
Although the forward model incorporates magnetic fields, the limited scale and largely bipolar distribution of magnetic elements are not direct analogues to all solar structures.
As reported in \citet{leenaarts_12}, the 3D radiation field (as opposed to a 1D vertical flux) can have a strong effect on the observed structure of the chromosphere.
Ultimately, the distribution of Mg II ions, the relative populations of the ground and doublet states, and the incident radiation field along the line of sight are the determining factors for the shape of the emergent Mg II h profile.
Both semi-empirical and MHD atmospheres provide us a means of calculating that distribution with strict limits on which physics can be included.\\
\indent The analysis of optically thick spectral lines present researchers with a double edge sword: the shape of the emergent profile encodes the thermal and hydrodynamic information of a significant vertical swath of the atmosphere, but the physical quantities we ultimately seek to measure are entangled.
One potential diagnostic is a Mg II profile inversion, similar to those used for optical spectropolarimetric data \citep{ruiz_92}. 
Of course inversions codes have largely operated in the regime where LTE is a reasonable approximation, which is not true for Mg II.
The next step in this research is the integration of our derived quantities into our physical models (cartoon or otherwise) of the how the dynamic chromosphere is structured.\\
\indent In this paper, we have dissected a dataset containing Mg II h spectral profiles over the full solar disk.
We decomposed the dataset into structural regions based on the magnetic field and quantified the differences between the regions based on a number of profile statistics.
The internetwork is dim at h2 and deeply reversed.
The network is brighter at h1, h2, and h3 than internetwork.
It has enhanced h1 width compared with internetwork.
Plage is brighter and more variable at h2 than network or internetwork.
It has a small h2 width and weak reversal.
Single peaked profiles occur all over the disk, but most commonly in plage.
We have created maps of many statistics for the line profiles.
There are coherent but varied structures visible in width, intensity, and velocity statistics.
Fibrils are primarily visible in plage in $\lambda_{h3}$ and in network in $I_{h3}$.
Both strong and weak field regions exhibit a great deal of profile variability which is the result of chromospheric dynamics.
We find strong correlations for the Doppler shift of the h3 core and the asymmetry in the h2 peaks and also between the widths of h1 and h2.
We calculated the center-to-limb variation of a number of statistics and compare them with the predictions of a semi-empirical model.
We find that both h3 and h2 exhibit limb darkening such that the core depth is flat as a function of $\mu$, while the semi-empirical model does not predict a strong limb darkening for h3.
Our measurement of this effect is most robust in internetwork.
This study presents the most comprehensive observational study on solar Mg II spectra to date, but it is clear that there are many facets on the formation and variability of this spectral line that are still not fully understood.
It is essential to keep developing atmospheric models which can accommodate the great deal of dynamical and structural variability we observe in high resolution observations of the chromosphere.
The photospheric velocity field strongly modulates the plasma conditions and radiation field in the lower chromosphere, and the upper chromosphere is highly structured by the magnetic field in the $\beta\le1$ regime.
Both of these boundary conditions must be considered in our future modeling efforts.

\end{document}